\documentclass{appolb}
\usepackage{graphicx}

\usepackage{amsmath}
\usepackage{amssymb}

\newcommand{\compresslist}{%
\setlength{\itemsep}{0pt}%
\setlength{\parskip}{0pt}%
\setlength{\parsep}{0pt}%
}

\begin{document}
\pagestyle{plain}

\title{MODEL OF COMMUNITIES ISOLATION AT HIERARCHICAL MODULAR NETWORKS}
\author{Pawe{\l} Kondratiuk and Janusz A. Ho{\l}yst
\address{Faculty of Physics, Center of Excellence for Complex Systems Research, Warsaw University of Technology, Koszykowa 75, PL-00-662 Warsaw, Poland}
}
\maketitle
\begin{abstract}
The model of community isolation was extended to the case when  individuals are randomly placed at nodes  of hierarchical modular networks. It was shown that the average number of blocked nodes (individuals) increases in time as a power function, with the exponent depending on  network parameters. The distribution of time when the first isolated cluster appears is unimodal, non-gaussian. The developed analytical approach is in a good agreement with the simulation data.
\end{abstract}


\section{INTRODUCTION}

Recently, hierarchical systems have been attracting attention of scientists working on complex networks \cite{Ravasz03,Komosa11,Laguna05,Galam05,Suchecki05Oscil}. In fact many real networks are hierarchically organized, e.g.  WWW network, actor network, or the semantic web \cite{Ravasz03}. Dynamics at such  networks can be qualitatively and quantitatively different from that at regular  lattices (see \cite{Komosa11,Laguna05,Galam05}).

The Ising model at a network with  a hierarchical topology was studied by Komosa and Ho{\l}yst \cite{Komosa11}. The analyzed parameters were, among others, magnetization, magnetic susceptibility, critical temperature and correlations of magnetization between different hierarchies. It was shown that the critical temperature is a power function of the network size and of the ratio $\frac{\langle k^2 \rangle}{\langle k \rangle}$, where $k$ stands for a node degree.

Opinion formation in hierarchical organizations was studied by Laguna et al. \cite{Laguna05}. Agents, belonging to various authority strata, try to influence  others opinions. The probability that an opinion of an agent of a certain authority prevails in the community depends on the size distribution of the authority strata. Phase diagrams can be obtained, where each phase corresponds to a distinct dominant stratum (or a sequence of the strata, with the decreasing probability of prevailing).

Fashion phenomena at hierarchical networks were studied by Galam and Vignes \cite{Galam05}. Interactions were imposed between social groups at different levels of hierarchy. A renormalization group approach was used to find the optimal investment level of the producer and to assess the influence of counterfeits on the probability of a new product  success.

One of fundamental topics in social dynamics are conflict situations and  many different sociophysics approaches \cite{Dornic01,Deffuant02,Galam02} or \emph{prisoner's dilemma}-type games \cite{Lee11} have been proposed. Recently a simple model of communities isolation has been introduced by Sienkiewicz and Ho{\l}yst \cite{Sienkiewicz09}. The model can describe such various issues as strategy at battlefields or formation of cultures. The idea behind this model is similar to the game of Go and it takes into account a natural leaning of people to avoid being surrounded by members of another (potentially hostile) community \cite{Schelling71}.

In this paper we extend the model of communities isolation  studied for chains, hypercubic, random and scale-free networks \cite{Sienkiewicz09,Sienkiewicz10}  to hierarchical networks proposed by \cite{Ravasz03}. 

\section{HIERARCHICAL NETWORKS}\label{sec:networks}

\begin{figure}[h]
\centering
\includegraphics[scale=0.5]{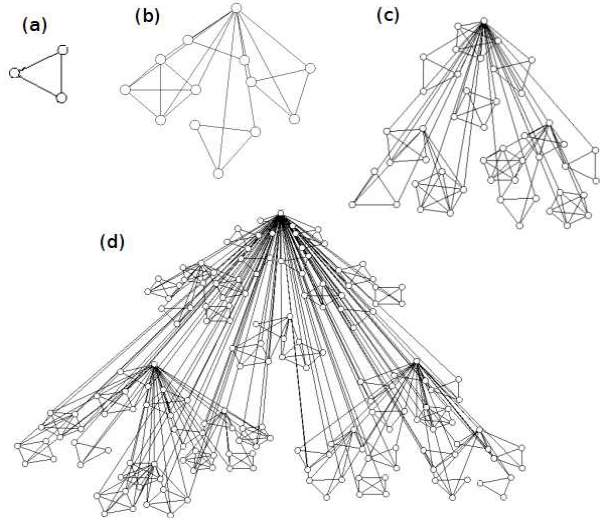}
\caption{Sample P1 networks with parameters $P_M(m) = \text{Unif}(2,4), p = 0.5$, with different degrees of hierarchy: (a) $h=0$, (b) $h=1$, (c) $h = 2$, (d) $h=3$}
\label{fig:sample_networks}
\end{figure}

The model of hierarchical networks was proposed by  Ravasz and Barab\'asi \cite{Ravasz03} and modified by Suchecki and Ho{\l}yst \cite{Suchecki05Oscil}. Such networks possess 3 parameters determining their structure:
\begin{itemize}
	\compresslist
	\item The degree of hierarchy $h \in \mathbb{N} \cup \{0\}$
	\item The distribution $P_M(m)$, where $m \in \mathbb{N}$, determining number of nodes at each level of hierarchy (in particular, the size of the cliques at the lowest level of hierarchy is $m+1$)
	\item The parameter determining the density of edges $p \in [0,1]$
\end{itemize} 
Two models (referred to as P1 and PD models) were analyzed, which differ in the density of edges. Each network has a central node, referred to as a \emph{center of hierarchy}. A network of hierarchy $h=0$ is a complete graph of size $m+1$ ($m$ is a random number, chosen with probability $P_M(m)$). The \emph{center of hierarchy}, due to the symmetry, is an arbitrary node. In order to construct a network of hierarchy $h>0$, one has to construct $m+1$ subnetworks of hierarchy $h-1$ and choose one of them --- its \emph{center of hierarchy} becomes a \emph{center of hierarchy} $v$ of the whole network. Afterwards, new connections (edges) are created: for each node $w$ of $m$ remaining subnetworks a connection (edge) $(v,w)$ is created with probability $p$ (in case of the P1 model) or $p^h$ (in case of the PD model). Sample networks created this way are presented in fig. \ref{fig:sample_networks}. Let us stress that the subnetworks do not have to be connected, especially if $p$ is small.

Some basic properties of such networks can be concluded from the construction algorithm:
\begin{itemize}
	\compresslist
	\item For $h \in \{0,1\}$, as well as for $p \in \{0, 1\}$, models P1 i PD are equivalent.
	\item For $p = 0$ the network consists of isolated cliques of size $m+1$ ($m$ --- random variable).
	\item For $P_M(m') = \delta_{m',m}$ the number of nodes (vertices) of the network equals $N = (m+1)^{h+1}$.
\end{itemize}
Periodic oscillations in degree distribution of such networks can be observed in log-log scale. The period, the amplitude and the shape of the peaks depend on the parameters of the network \cite{Suchecki05Oscil}.

In this paper only the case with $P_M(m') = \delta_{m',m}$ ($m=const$) was considered, which corresponds to the original Ravasz and Barab\'asi model \cite{Ravasz03}.


\section{BASIC ISOLATION MODEL}\label{sec:basic_model}

The model of communities isolation was proposed by Sienkiewicz and Ho{\l}yst \cite{Sienkiewicz09}. The rules are similar to those of the game of Go. A number of communities compete with each other, settling nodes of a network. In each step a random empty node is chosen. It is then settled by a member of randomly chosen community. A cluster of nodes occupied by one community becomes blocked when it gets surrounded by another community. The surrounded nodes are no more active in the game, i.e. they can not take part in surrounding other communities.   

The case of communities competing at a chain was analyzed in \cite{Sienkiewicz09}. Two functions describing the evolution were studied: the average number of blocked nodes over time and a mean critical time, i.e. the moment, when the first blocked cluster appears. In \cite {Sienkiewicz10} the influence of external bias was considered when  settling rates of competing communities are different.

In this paper the case of two competing communities at P1 and PD hierarchical networks is considered. Two parameters are analyzed: the average number of blocked nodes $Z(t)$ and the critical time distribution $Pr(t_c)$.


\section{NUMBER OF BLOCKED NODES OVER TIME}\label{sec:Z}

\subsection{Case $p=0$}

For $p = 0$ models P1 are PD equivalent. The network consists of $\frac{N}{m+1}$ isolated cliques of size $m+1$. In such case

\begin{equation}
  \left\{
  \begin{array}{l l}
  Z(0)  =  0 \\
  Z(t+1)  =  Z(t) + \sum_{i=1}^{m} i p_i,
  \end{array} \right.
\end{equation}
where $p_i$ --- probability that in the $(t+1)$th step $i$ nodes will be blocked,
\begin{equation}
p_i = \left(\frac{t}{2N}\right)^m \binom{m}{i}.
\end{equation}
After short algebra we obtain
\begin{equation}
\left\{
\begin{array}{l l}
Z(0)  =  0 \\
Z(t+1) = Z(t) + \frac{m}{2} \left(\frac{t}{N}\right)^m.
\end{array} \right.
\end{equation}
The solution of this recursive equation is a $(m+1)$th degree polynomial, which can be approximated by substituting the sum with the integral:
\begin{eqnarray}
Z(t) & = & \sum_{i=0}^{t-1} \frac{m}{2} \left(\frac{i}{N}\right)^m \approx \int_{0}^{t} \frac{m}{2} \left(\frac{x}{N}\right)^m dx \nonumber\\
& = & \frac{m}{m+1} \frac{t^{m+1}}{2 N^m}.
\label{eqn:Z_p0}
\end{eqnarray}

As one can see, $Z(t)$ is a power function. The exponent $\beta$ depends only on the $m$ parameter, $\beta=m+1$.

\subsection{Case $p=1$}

In this case models P1 and PD are also equivalent.
For networks of hierarchy $h=1$:
\begin{eqnarray}
Z^{(1)}(t) & = & \rho_0 \left( \rho_1^m \frac{m}{2} + m \rho_1^{m+1} \frac{m+1}{2} \right) \nonumber\\
& = & \frac{1}{2} m \rho_0 \rho_1^m (1 + (m+1) \rho_1 ),
\end{eqnarray}
where $\rho_i$ is a \emph{reduced density}:
\begin{equation}
\rho_i = \rho_i(t) \equiv \left\{
		\begin{array}{l l}
		0 & \quad \text{for $t < i$}\\
		\frac{t-i}{N-i} & \quad \text{for $t \geq i $}\\
		\end{array} \right.
\end{equation}

\begin{figure}[h]
\centering
\includegraphics[scale=0.48]{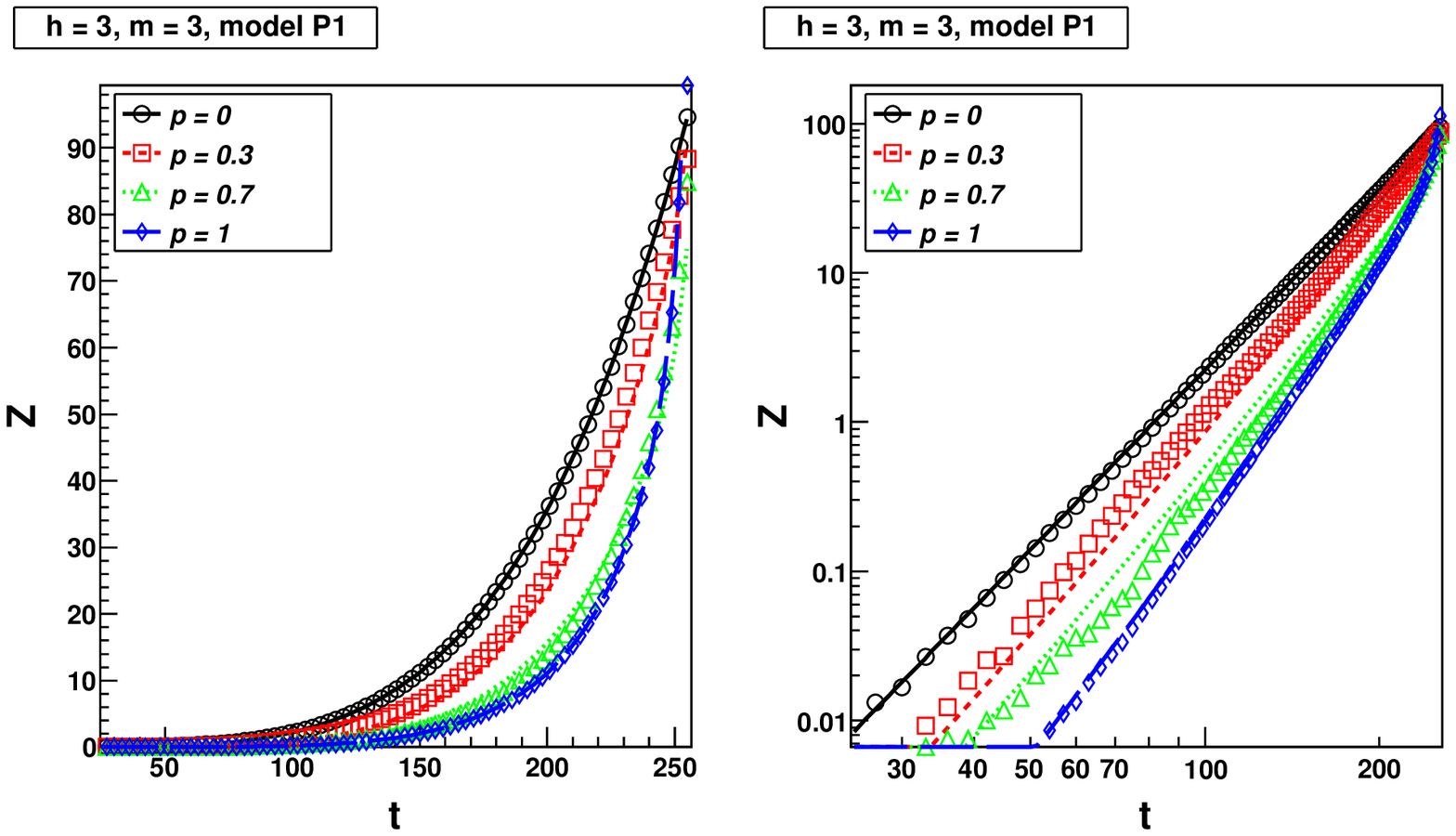}
\includegraphics[scale=0.48]{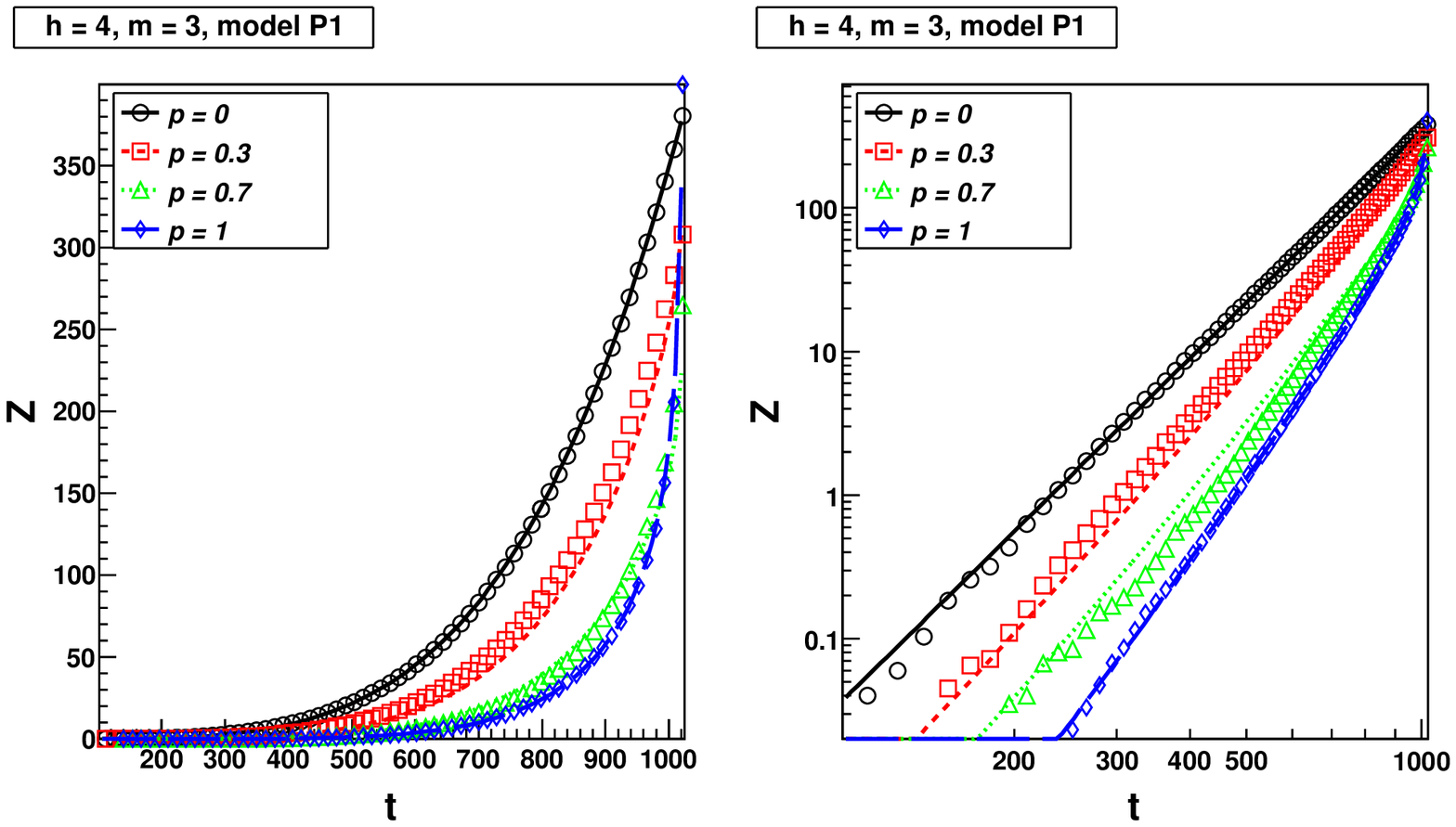}
\includegraphics[scale=0.48]{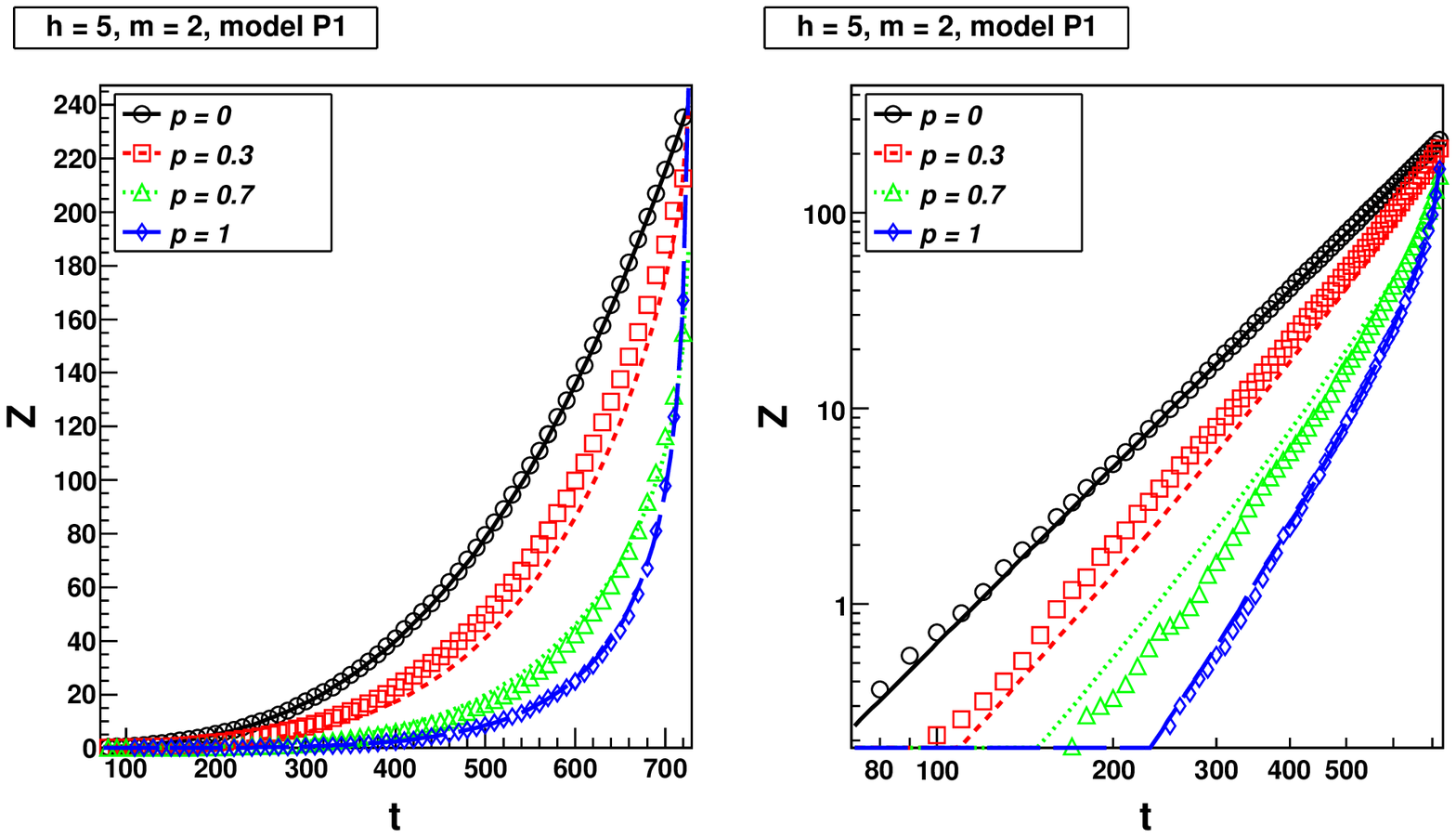}
\caption{(color online) Average number of blocked nodes, $Z(t)$, for various networks of P1 model. Symbols correspond to data from computer simulations. Lines show analytical approximations (eq. \eqref{eqn:Z_p0}, \eqref{eqn:Z_p1} and \eqref{eqn:Z_general}). Left side --- linear scale, right side --- log-log scale.}
\label{fig:Z_P1}
\end{figure}

\begin{figure}[h]
\centering
\includegraphics[scale=0.48]{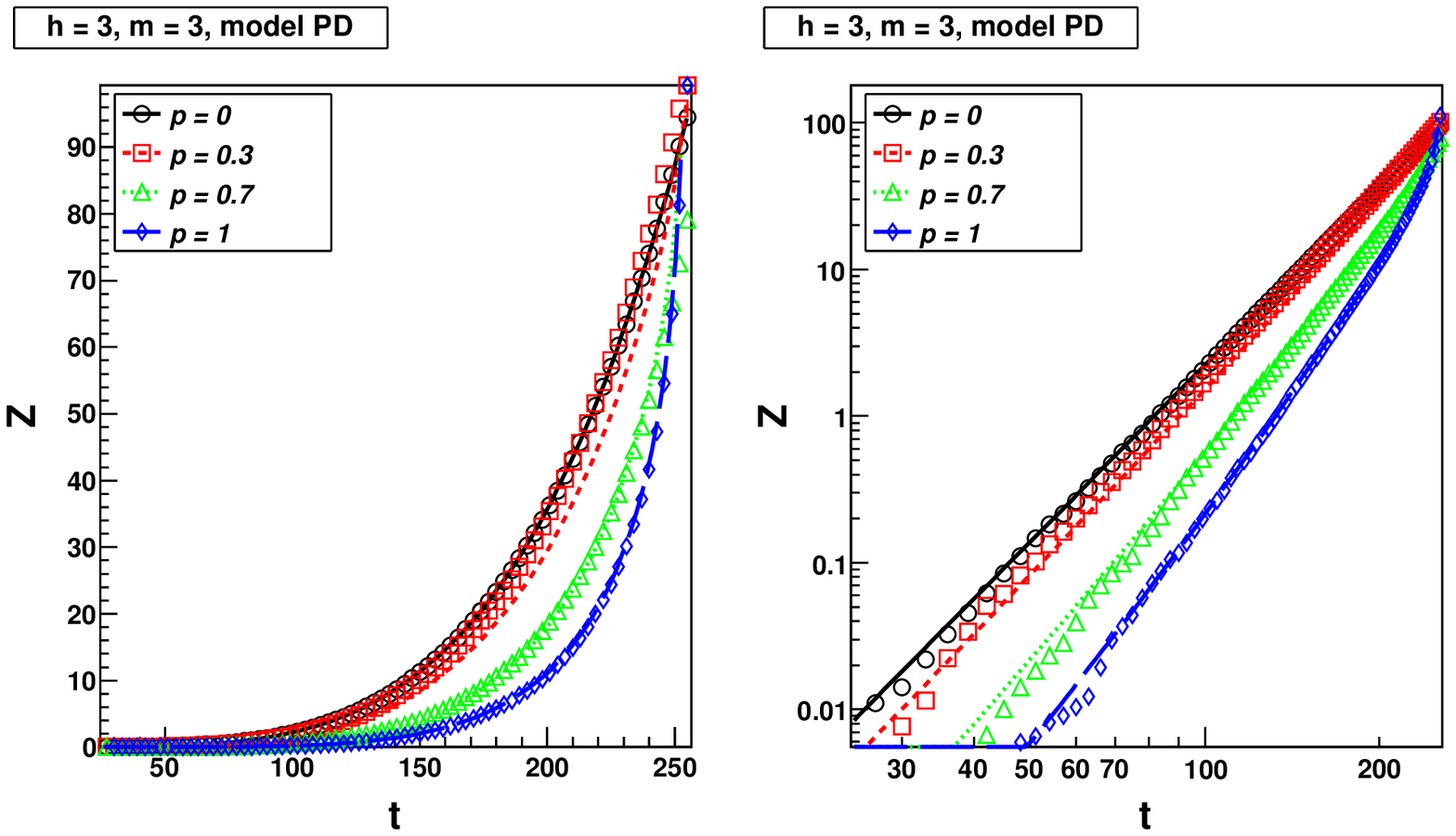}
\includegraphics[scale=0.48]{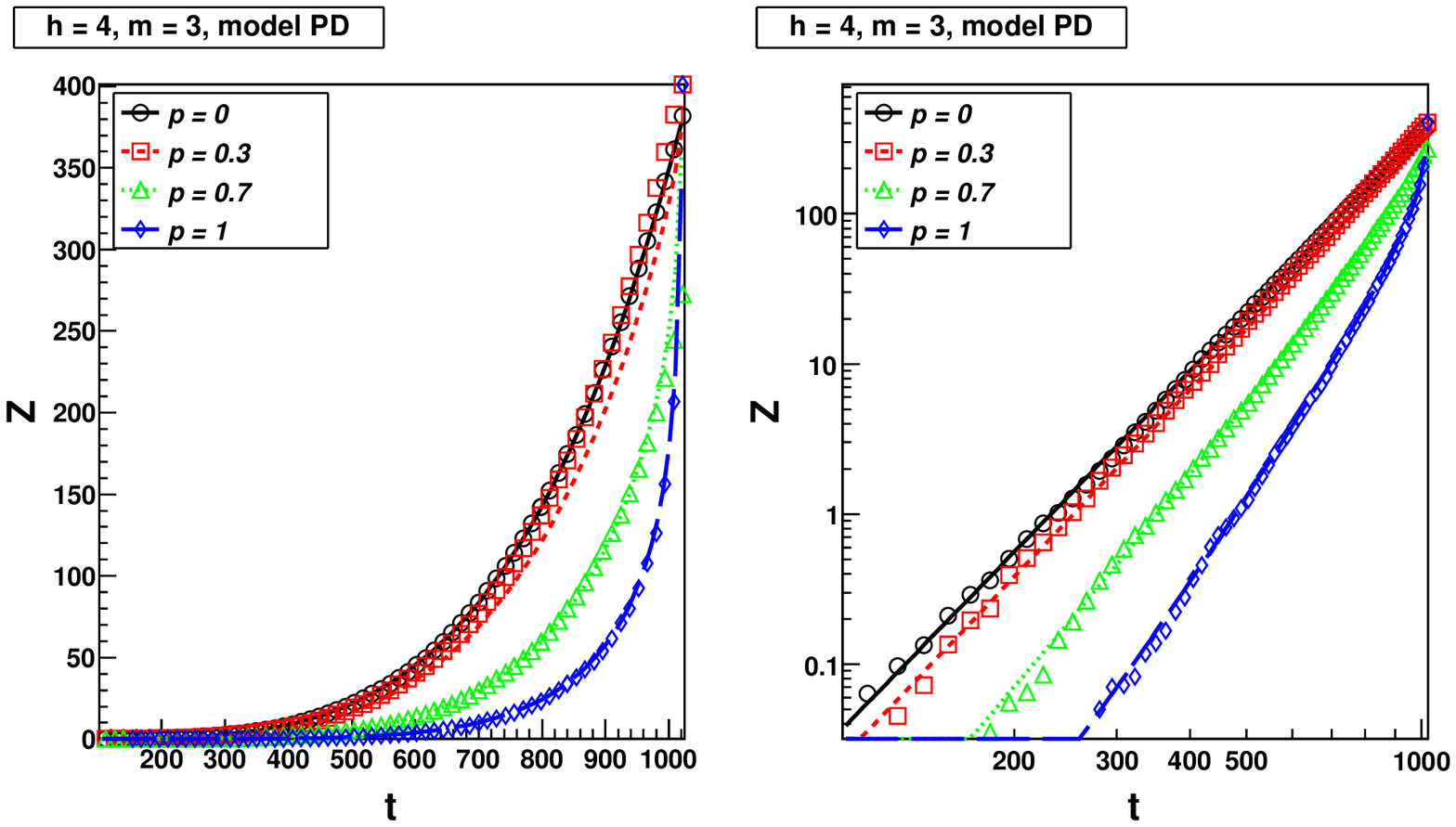}
\includegraphics[scale=0.48]{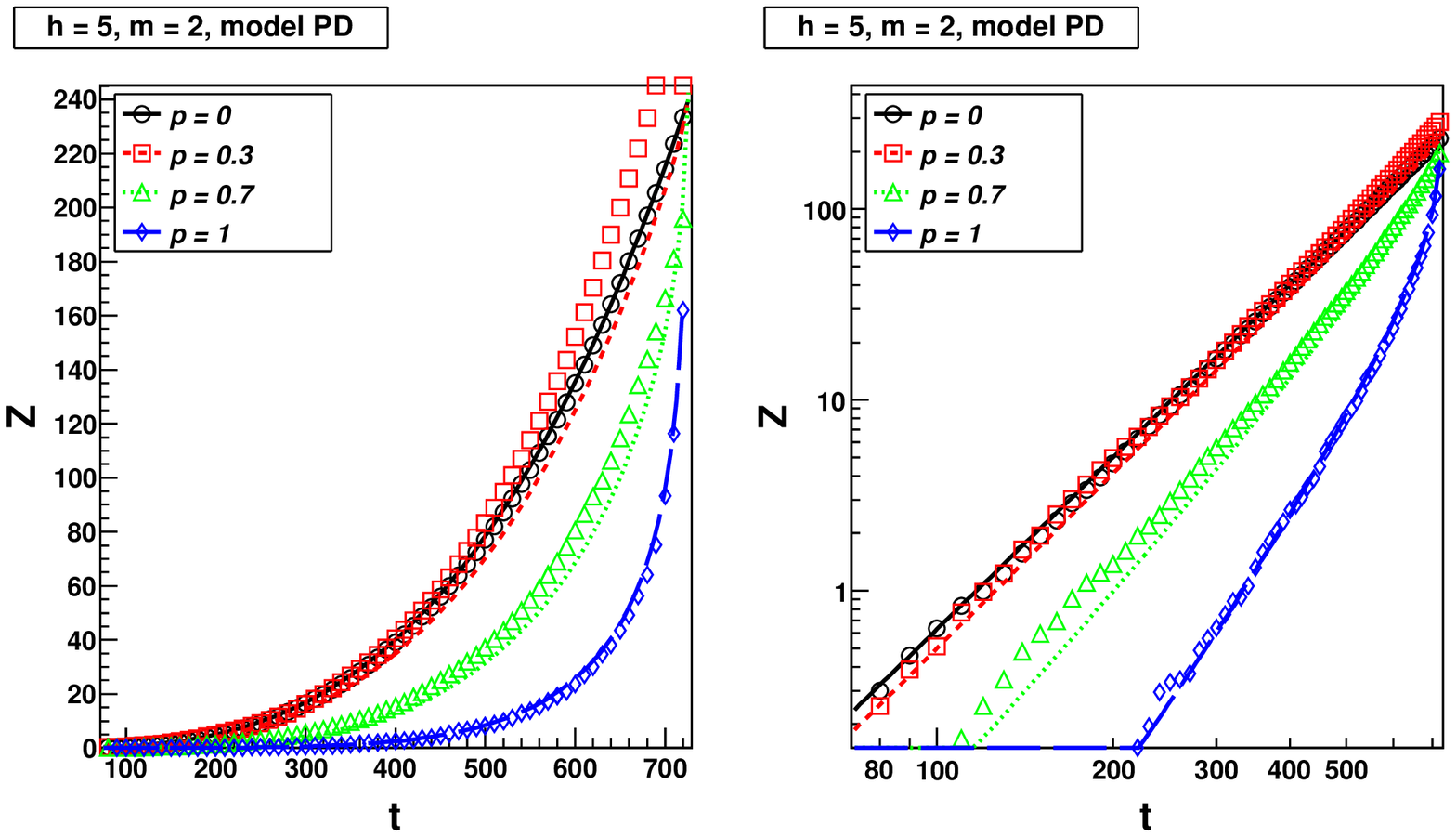}
\caption{(color online) Average number of blocked nodes, $Z(t)$, for various networks of PD model. Symbols correspond to data from computer simulations. Lines show analytical approximations (eq. \eqref{eqn:Z_p0}, \eqref{eqn:Z_p1} and \eqref{eqn:Z_general}). Left side --- linear scale, right side --- log-log scale.}
\label{fig:Z_PD}
\end{figure}

For networks of higher hierarchies, $h \geq 1$, a recursive equation well approximating $Z^{(h)}(t)$ can be derived. The idea behind the formulas is that a clique can only be blocked if all the nodes of higher hierarchies neighboring with it are filled. Therefore $Z^{(h)}(t) = 0$ if the center of hierarchy of the network (which neighbors with all the other nodes) is empty. In the opposite case, $Z^{(h)}(t)$ depends on $Z^{(h-1)}(t)$, which describes each of $m+1$ \emph{subnetworks}.
\begin{equation}
\left\{
\begin{array}{l c l}
Z_i^{(1)}(t) & = & \frac{1}{2} m \rho_i \rho_{i+1}^m \left(1 + (m+1) \rho_{i+1}\right) \\
Z_i^{(h)}(t) & = & Z_i^{(h-1)}(t) + \frac{1}{2} m \rho_i Z_{i+1}^{(h-1)}(t) \\
& + & \frac{1}{4} \rho_i \rho_{i+1} \rho_{i+2}^{(m+1)^h-1} \left( (m+1)^h + 1 \right) \\
Z^{(h)}(t) & \equiv & Z_0^{(h)}(t)
\end{array}
\right.
\label{eqn:Z_p1}
\end{equation}
This equation can only be solved numerically. The solutions are presented in fig. \ref{fig:Z_P1} and \ref{fig:Z_PD}. It can be noticed, that within a wide range of time $t$, $Z^{(h)}$ can be with reasonable accuracy approximated with a power function
\begin{equation}
Z^{(h)}(t) \propto t^\beta,
\end{equation}
where the exponents $\beta$ are higher than in the case of $p=0$ and they are close to $m+4$.

\subsection{General case}

An analytical approximation of $Z^{(h)}(t)$ for networks with higher hierarchies ($h>1$) when the parameter $p$ is different from zero and one is far more difficult. Instead of searching for such a formula, an alternative approach was chosen. It was assumed that $Z^{(h)}(t)$ can be estimated from the proportion
\begin{equation}
\frac{\log Z_{p=0}^{(h)}(t) - \log Z^{(h)}(t)}{\log Z^{(h)}(t) - \log Z_{p=1}^{(h)}(t)} \approx \frac{f(p,h)}{1 - f(p,h)},
\label{eqn:Z_general}
\end{equation}
where $f(p,h) \in [0,1]$ should be an increasing function of $p$ which, while not being too complicated, would give a reasonable approximation for the widest possible ranges of $p$ and $h$. It turned out that in the case of the P1 model, choosing $f(p,h) = p$ results in a good agreement of the $Z^{(h)}(t)$ function with simulation  data. For the PD model, $f(p,h) = p^{\frac{h}{2}}$ is a good choice.

\section{CRITICAL TIME DISTRIBUTION}\label{sec:tc}

\subsection{Case $p=0$}

As it was previously mentioned, in the case of $p=0$ the network consists of $\frac{N}{m+1} = (m+1)^h$ isolated cliques of $m+1$ nodes. In order to find the distribution of critical time (time, when the first blocked cluster appears), one has to consider the probability that at time $t$ there are no blocked nodes yet. It means that at time $t$ the only completely filled cliques are those filled with members of one community, which leads to the formula
\begin{eqnarray}
Pr(t_c > t) = \left(1 - \alpha \left(\frac{t}{N}\right)^{m+1} \right)^\frac{N}{m+1},
\end{eqnarray}
where $\alpha \equiv 1-2^{-m}$. The cumulative critical time distribution can be immediately obtained
\begin{equation}
Pr(t_c \leq t) 
= 1 - \left(1 - \alpha \left(\frac{t}{N}\right)^{m+1} \right)^\frac{N}{m+1},
\end{equation}
as well as the critical time distribution in the approximation of continuous time:
\begin{eqnarray}
\label{eqn:tcDist_p0}
Pr(t_c = t) & = & Pr(t_c \leq t) - Pr(t_c \leq t-1) \nonumber\\
& \approx & \frac{d}{dt} Pr(t_c \leq t) \nonumber\\
& = & \alpha \left(1 - \alpha \left(\frac{t}{N}\right)^{m+1} \right)^{\frac{N}{m+1} - 1}  \left(\frac{t}{N} \right)^m.
\end{eqnarray}

The mean critical time can be also calculated analytically:
\begin{eqnarray}
\label{eqn:avgTc_p0}
\langle t_c \rangle & = & \int_0^N t Pr(t_c = t) dt \approx \int_0^N t \frac{d}{dt} Pr(t_c \leq t) dt  \nonumber \\
& = & \frac{N}{m+1} \alpha^{-\frac{1}{m+1}} \left( B\left(\frac{N}{m+1} + 1,\frac{1}{m+1}\right) \right. \nonumber\\
& & - \left.B\left(2^{-m};\frac{N}{m+1} + 1,\frac{1}{m+1}\right) \right),
\end{eqnarray}
where $B(a,b) \equiv \int_0^1t^{a-1}(1-t)^{b-1}\,dt$ (Euler beta function) and $B(x;a,b) \equiv \int_0^xt^{a-1}(1-t)^{b-1}\,dt$ (incomplete Euler beta function).

\subsection{Case $p=1$}

For networks of hierarchy $h=0$
\begin{equation}
Pr(t_c > t) = 1 - \alpha \left( \frac{t}{N} \right)^{m+1} = 1 - \alpha \rho_0^{m+1}.
\end{equation}
For networks with hierarchy $h=1$
\begin{equation}
Pr(t_c > t) = 1 - \rho_0 + \rho_0 (1 - \rho_1^m) \left( 1 - \alpha \rho_1^{m+1} \right)^m.
\end{equation}
For networks with any degree of hierarchy, $h \geq 0$, a recursive formula for the cumulative critical time distribution can be expressed as
\begin{equation}
\label{eqn:tcDist_p1}
\left\{
\begin{array}{l}
F_i^{(0)}(t) = \alpha \rho_i^{m+1} \\
F_i^{(h)}(t) = \rho_i - \rho_i (1 - \rho_{i+1}^m) \left( \prod\limits_{d=0}^{h-1} (1 - F_{i+1}^{(d)}(t)) \right)^m \\
Pr^{(h)}(t_c \leq t) \equiv F_0^{(h)}(t).
\end{array}
\right.
\end{equation}

\begin{figure}[h]
\centering
\begin{tabular}{cc}
\includegraphics[scale=0.28]{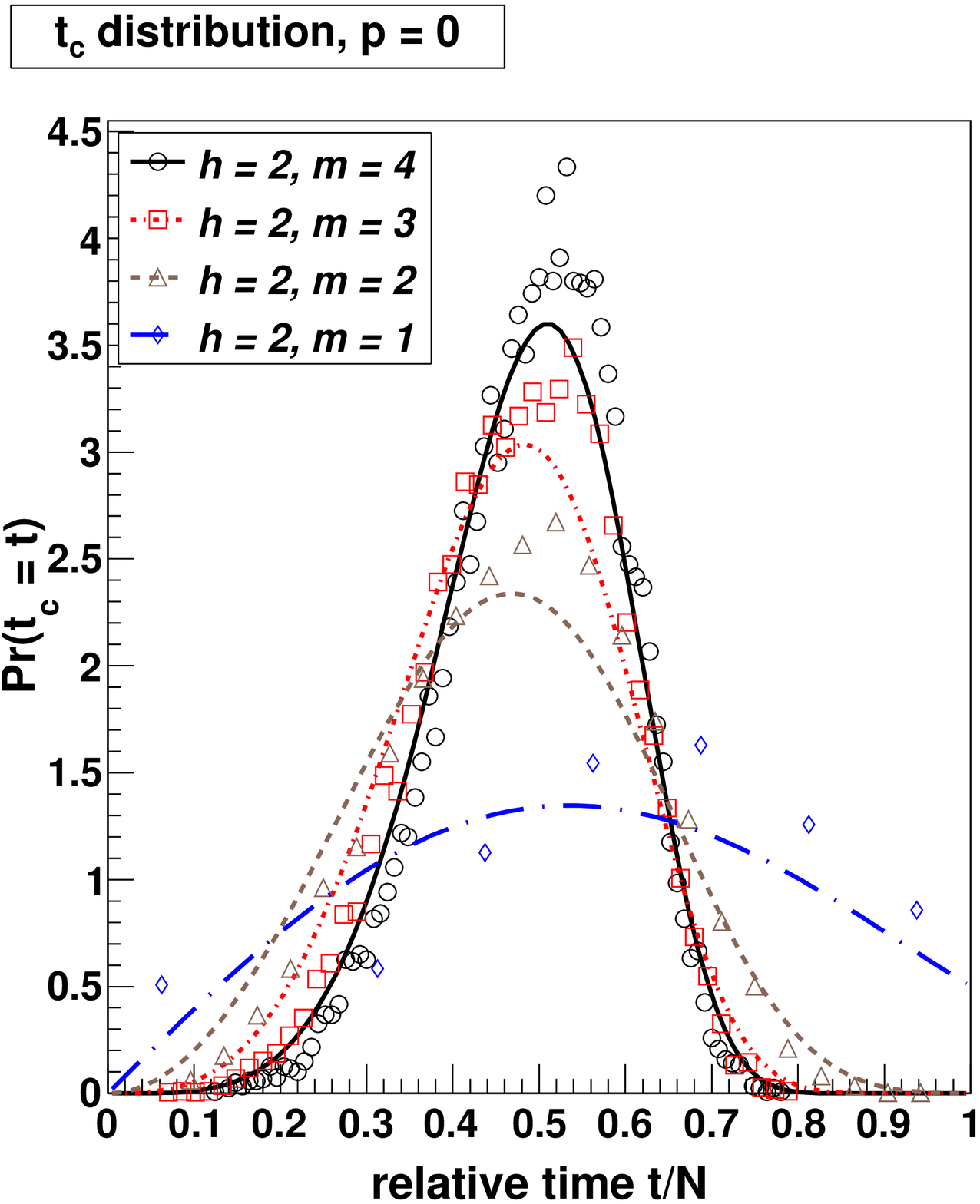} & \includegraphics[scale=0.28]{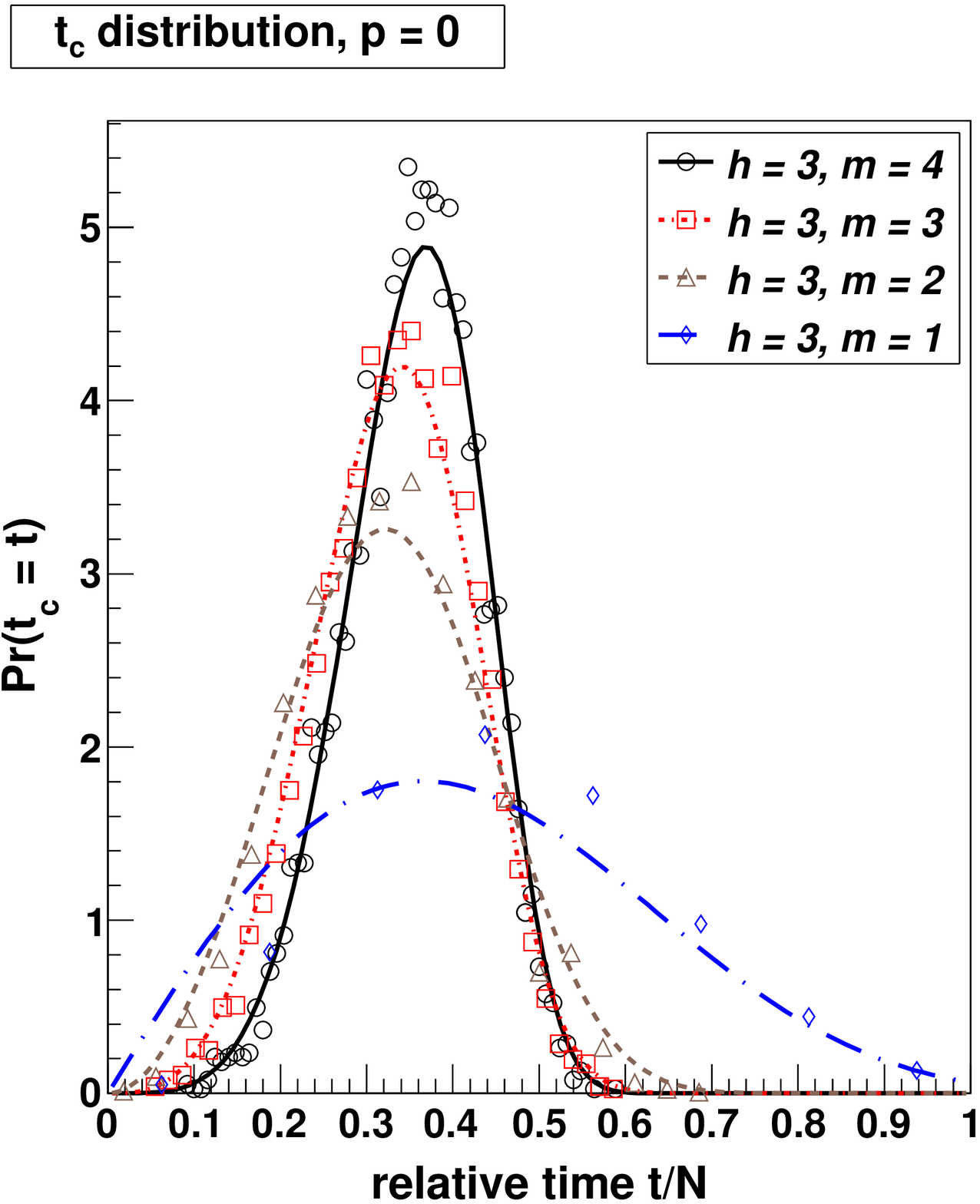} \\
\includegraphics[scale=0.28]{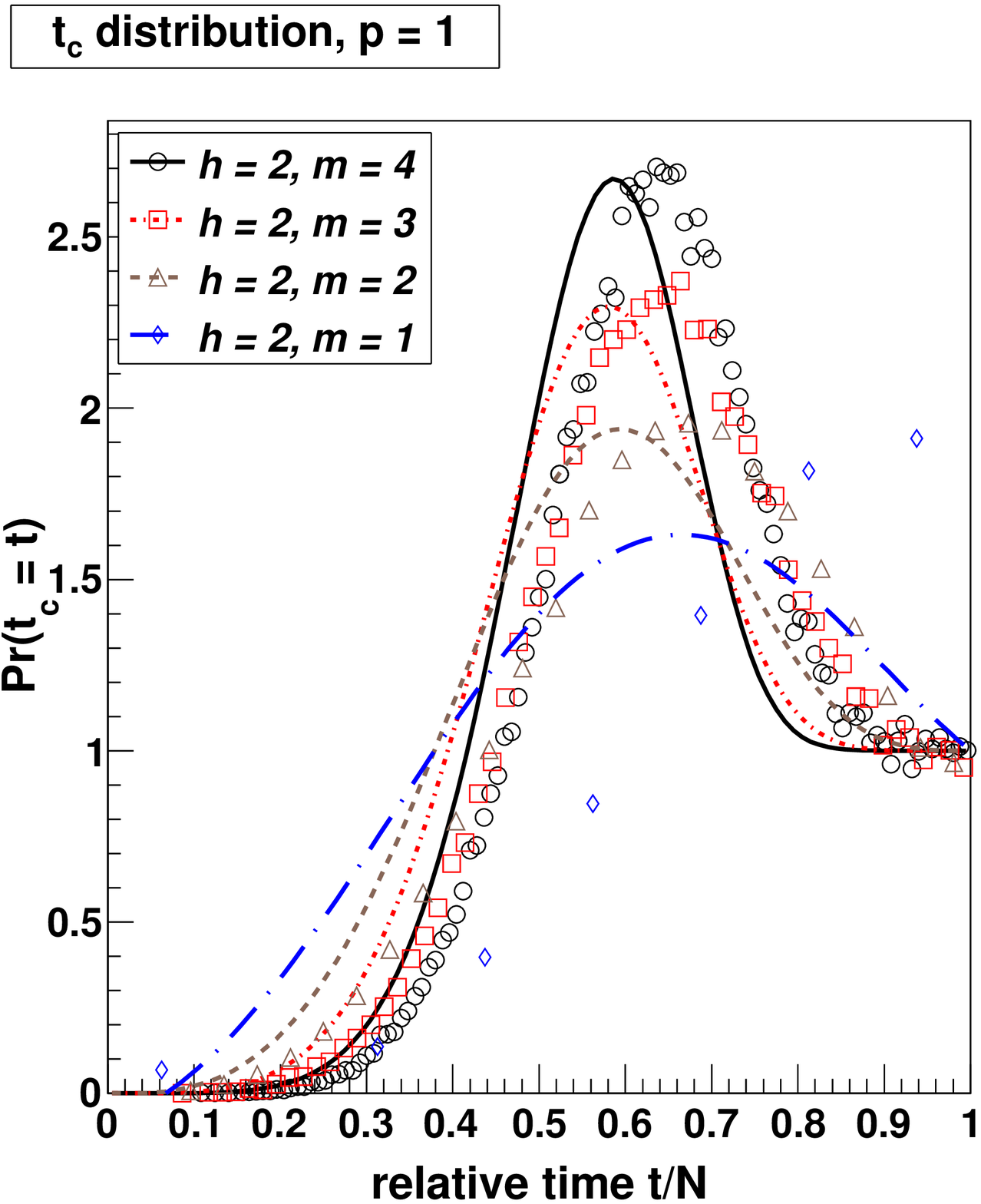} & \includegraphics[scale=0.28]{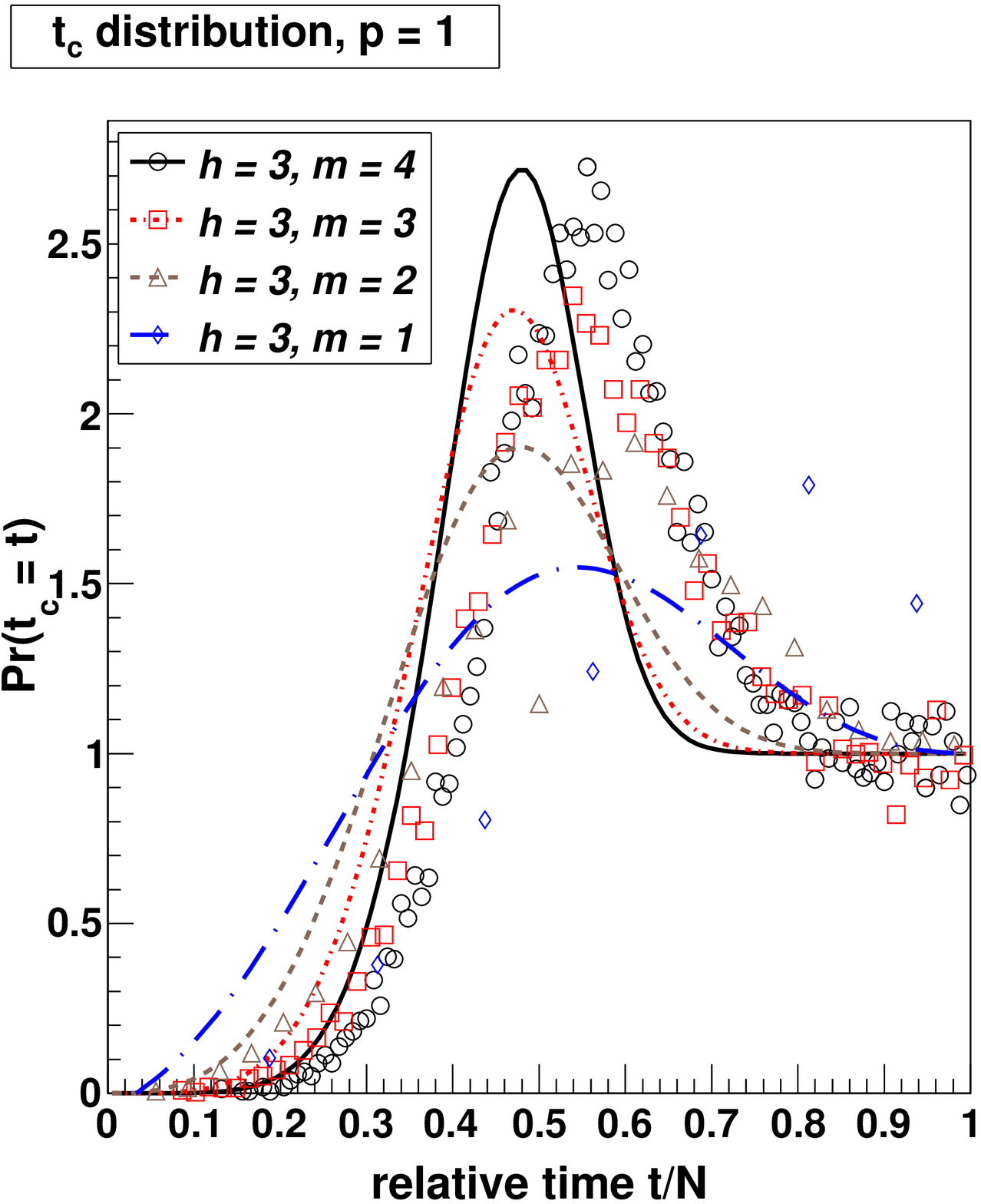} \\ 
\end{tabular}
\caption{(color online) Critical time distribution for networks of hierarchy $h=2$ (left) and $h=3$ (right), with $p=0$ (top) and $p=1$ (bottom). Symbols --- simulated data, smooth lines --- analytical approximation (eq. \eqref{eqn:tcDist_p0} and \eqref{eqn:tcDist_p1}).}
\label{fig:tc}
\end{figure}

The mean critical time can be obtained by numerical integration of $Pr^{(h)}(t_c \leq t)$:
\begin{equation}
\label{eqn:avgTc_p1}
\langle t_c \rangle = N - \int_0^N Pr^{(h)}(t_c \leq t) dt.
\end{equation}

\begin{figure}[h]
\centering
\includegraphics[scale=0.4]{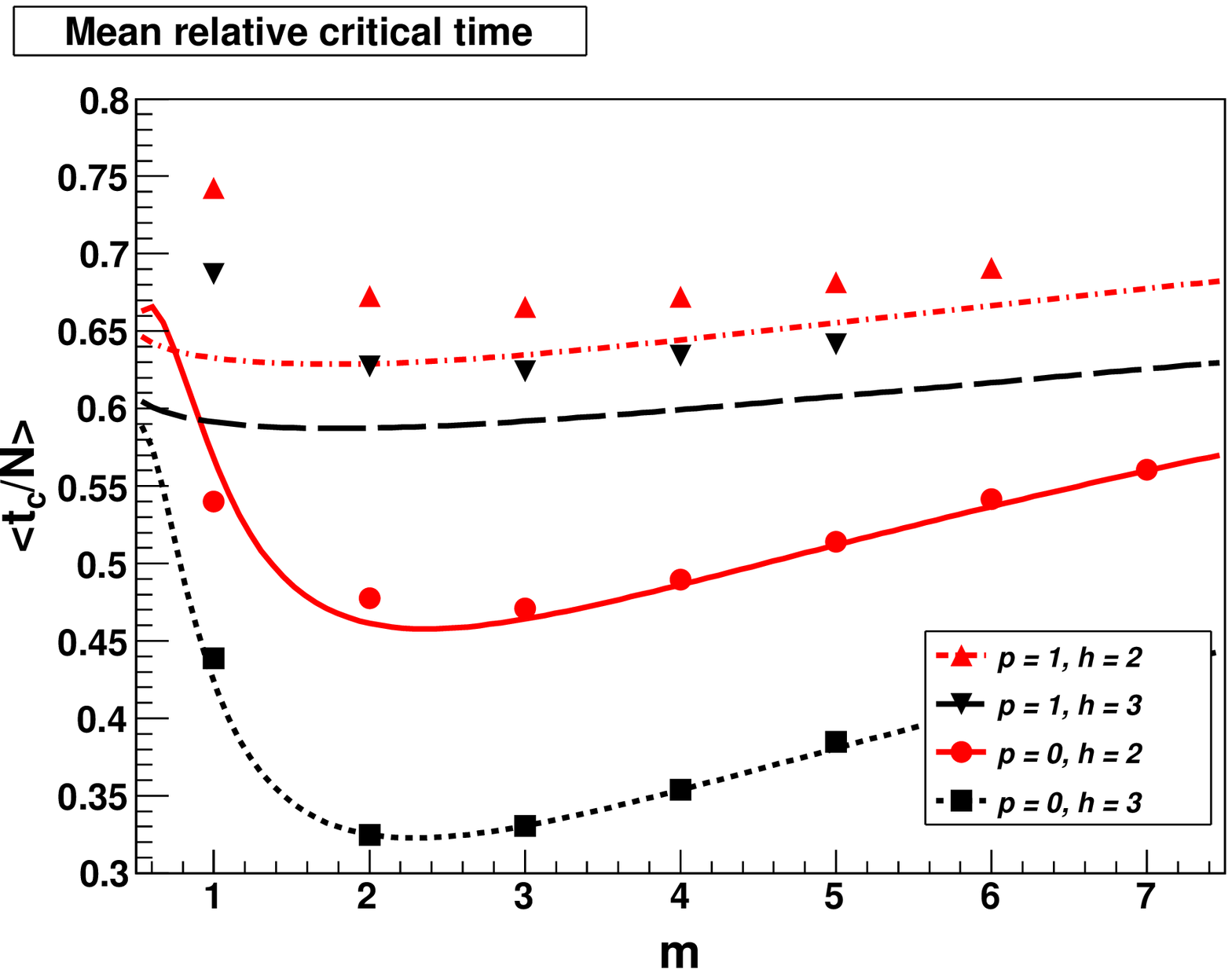}
\caption{(color online) Mean critical time for various networks. Symbols correspond to data from the simulations, lines --- to the analytical approximations (eq. \eqref{eqn:avgTc_p0} and \eqref{eqn:avgTc_p1}).}
\label{fig:mean_tc}
\end{figure}


\section{DISCUSSION AND CONCLUSIONS}
\label{sec:disc_conc}

\subsection{Number of blocked nodes $Z(t)$}

In all cases the  function $Z(t)$, defined as the average number of blocked nodes at time $t$, can be approximated with a high accuracy by a power function
\begin{equation}
Z(t) \propto t^\beta,
\end{equation}
The $\beta$ exponent depends on the parameters of the network. For $d$-dimensional hypercubic networks (including the $1$-dimensional ones, i.e. chains) $\beta = 2d-1$. For modular hierarchical networks $\beta$ depends mainly on $m$ and $p$ parameters, i.e. on the sizes of basic cliques at the lowest hierarchy  and on the density of inter-clique connections. The dependence on the degree of hierarchy $h$ (and on the network size) is weak, what can be explained by the fact, that increasing the degree of hierarchy $h$ is a process similar to system rescaling, therefore
\begin{equation}
Z^{(h+1)}(\rho) \approx (m+1) Z^{(h)}(\rho) = (m+1) C \rho^\beta = C' t^\beta.
\end{equation}

For $p=0$, the parameter $\beta$ can be analytically found: $\beta = m+1$. The result is in agreement with the simulated data. Increasing the density of connections (the $p$ parameter) leads to the increase of $\beta$ --- up to approximately $m+4$ for $p=1$.

There is an important distinction in the way $Z(t)$ was approximated for hypercubic and hierarchical networks. For hypercubic networks, the number of isolated nodes was calculated using the following approximation: all  blocked nodes were blocked alone, i.e. they do not neighbor with other blocked nodes (of the same community). Although this approximation might seem coarse,  resulting analytical predictions turned out to be in quite good agreement with simulated data \cite{Sienkiewicz09,Sienkiewicz10}. For modular hierarchical networks, such an approximation would not be reasonable. Because of the fact that at the lowest level of hierarchy such networks consist of cliques of $m+1$ nodes, the most probable are situations when $\frac{m+1}{2}$ nodes are simultaneously blocked.

\subsection{Critical time $t_c$}

The second analyzed parameter was critical time $t_c$, i.e. the moment, when the first isolated cluster appears. It is a random variable. The critical time distribution $Pr(t_c)$ was studied, as well as mean critical time $\langle t_c \rangle$. More precisely, \emph{a critical density} (or \emph{a critical relative time})
\begin{equation}
\rho_c \equiv \frac{t_c}{N}
\end{equation}
was often shown  so networks with different parameters could be easily compared.

The $Pr(\rho_c)$ distribution is always unimodal. The mode ($\operatorname{arg\,max} Pr(\rho_c)$) decreases with the increase of $h$ and  the standard deviation $\sigma(\rho_c)$ decreases with $m$.

For $p=0$ it was possible to find the analytical formula for both $Pr(t_c)$ and $\langle t_c \rangle$. The distribution $Pr(t_c)$ is a polynomial of degree $m(m+1)((m+1)^h-1)$ (see eq. \ref{eqn:tcDist_p0}) and the average $\langle t_c \rangle$ is a scaled difference of two Euler beta functions (see eq. \ref{eqn:avgTc_p0}). The average $\langle \rho_c \rangle$ decreases with $h$ and for  a fixed $h$ it reaches a minimum for  $m\approx 2$ (see fig. \ref{fig:mean_tc}).

For $p=1$ the  distribution $Pr(\rho_c)$ reaches a constant, non-zero value for $\rho_c \in [1-\epsilon,1]$ (for $h \leq 3$, $\epsilon \approx 0.1$), which means that processes when blocked clusters firstly appear at the very end of the evolution are not unlikely.

The values of $\langle \rho_c \rangle$ can be compared with those obtained for hypercubic networks. Similar trends can be observed in hypercubic and hierarchical networks: $\langle \rho_c \rangle$ decreases with the network size $N$ and increases with the average degree. However, for modular hierarchical networks the dependence of $\langle \rho_c \rangle$ on the average degree (which equals $m$ for $p=0$ and rises with $p$) is very weak in comparison to hypercubic networks. Typical values of $\langle \rho_c \rangle$ for hierarchical networks correspond to the ones obtained for two- or three-dimensional networks, even for $m\gg 3$.

\begin{center}
{\bf Acknowledgments}
\end{center}

The  authors  acknowledge  support  from  the  European COST Action MP0801 {\it Physics of Competition and Conflicts} and from the Polish Ministry of Science Grant No. 578/N-COST/2009/0.


\end{document}